\documentclass[doublecol]{epl2}
\usepackage{amssymb,amsmath,latexsym}
\usepackage{epsfig}
\newcommand{\C}[1]{{\mathcal{#1}}}    
\begin{document}

\title{Exposing the static scale of the glass
transition by random pinning}
\author{Smarajit Karmakar and Itamar Procaccia}

\institute{Dept. of Chemical Physics, The Weizmann Institute of
Sceince, Rehovot 76100, Israel}

\abstract{
The dramatic slowing down associated with the glass transition cannot be fully understood without an associated
static length that is expected to increase rapidly as the temperature is reduced. The search for such a length
was long and arduous, without a universally accepted candidate at hand. Recently a natural such length $\xi_s$
was proposed, stemming from a cross-over between plastic and elastic mechanical responses of the material.
In this Letter we show that supercooled liquids in which there exists random pinning sites of density
$\rho_{\rm im}\sim 1/\xi_s^d$ exhibit complete jamming of all dynamics. This is a direct
demonstration that the proposed length scale is indeed {\em the} static length that was long sought-after.
}
\maketitle
A well accepted theory to explain the dramatic slowing down of the dynamics of super-cooled liquids upon approaching the glass transition is still  missing, even after decades of intense research efforts ~\cite{01Donth, 01DS}. One of the prominent
approaches draws an analogy to the slowing down near continuous phase transitions; such an analogy requires a growing length scale that can accommodate the dynamical slowing down.
The discovery of dynamic heterogeneity  in super-cooled
liquids both in experiments ~\cite{00Ediger} and in theoretical studies ~\cite{05Chi4,95HH} and the detailed
analysis of the associated dynamical length-scale using multi-point correlation functions
~\cite{99BDBG,02DFGP,00FP, 09KDS,10KDS} seemed to point in the right direction. But in Ref.
~\cite{09KDS} it was shown that the growth of the dynamical length-scale associated with the dynamic
heterogeneity upon decreasing the temperature is {\em not} directly related to the rapid increase in relaxation
time. It was also shown in Ref. \cite{09KDS} that the increase in relaxation time with decreasing temperature
is better correlated with the decrease in configurational entropy via the Adam-Gibbs relation
~\cite{65AG} which relates the structural relaxation time $\tau_{\alpha}$ to the configurational entropy $S_c$ as,
\begin{equation}
\tau_{\alpha} = \tau_0 \exp\left(\frac{A}{TS_c}\right)\ .
\label{AdamGibbs}
\end{equation}
Here $A$ is a temperature independent constant. This phenomenological relation is found to fit the data quite well
over a large range of relaxation time in many experiments. 

The question of the existence of a {\em static} length that
\begin{figure*}
\hspace{-0.5cm}
\includegraphics[scale = 0.50]{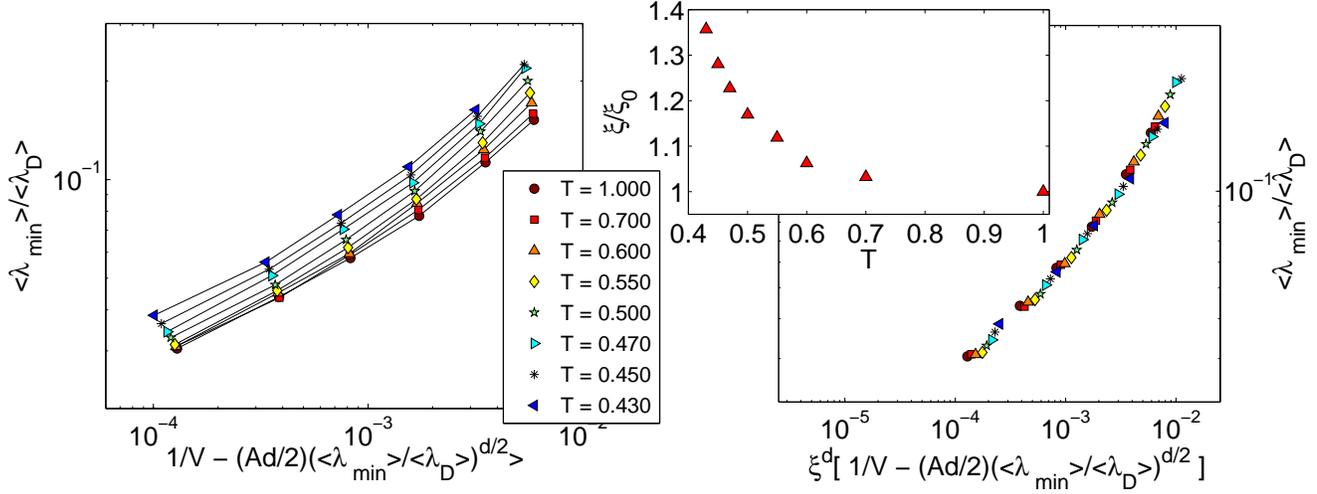}
\caption{(Color online) Left panel : The measured minimal eigenvalue as a function of system size for a
binary system interacting via Kob-Andersen potential in 3 dimension. Every data point represents an
average over 2000 Inherent structures which are the local energy minimum of the potential energy function.
Right panel : Same data set but plotted according to the scaling
ansatz Eq.~\ref{ansatz} to obtain the static length-scale. Inset shows the temperature dependence of the
obtained length-scale.}
\label{lamminKA}
\end{figure*}
characterizes the degree of disorder in the approached glassy state remained open.
Recently we were able to define and measure such a {\em static} length-scale $\xi_s$ \cite{11KLP}
that grows upon approaching the glass transition. To be sure, there are other candidates in the recent
literature, e.g a length-scale from the effects of boundary conditions, \cite{08BBCGV}, point to set
correlations \cite{07FM}, the scaling of the non affine displacement field \cite{10MGIO} and
patch correlation scale \cite{09KL}. In this Letter we present strong arguments to demonstrate
that the length-scale extracted by us via a different approach ~\cite{11KLP} is {\em the} relevant
static length-scale in determining the dynamics of the super-cooled liquids.

The idea to single out the relevant length scale is based on introducing by hand another length scale,
to observe the crossover phenomena expected in a system with two competing length-scales. If one has
two length scales that affect a relaxation process then the dominant one will be always the smaller
between the two. Thus by tuning the artificial length scale we can cross over from dynamics governed
by one length-scale to the other. A simple way to create a tunable length-scale is to introduce a
density $\rho_{im}$ of immobile paticles. The length-scale associated with this quenched disorder
will be $\xi_{im} \sim \rho^{-1/d}_{im}$, where $d$ is the space dimension. If our proposed length-scale
is indeed {\em the} static length scale then we expect to see a crossover from dynamics governed by $\xi_s$ to
a dynamics governed by the quenched disorder length-scale $\xi_{im}$.

Before going into the details of the numerical analysis we should briefly mention how to extract the
static length-scale $\xi_s$ (details can be found in \cite{11KLP}). Our starting point is the fact that
at low frequency tail of the density of state (DOS) of amorphous solids consisting of $N$ particles reflects the excess of plastic modes
which do not exist in the density of states of purely elastic solid. \cite{02TWLB,10Sok}. This excess of modes is sometime referred to as the `Boson Peak' \cite{09IPRS}.
Here and below the `mode' refers to the eigenfunction of the underlying Hessian matrix. Recently \cite{11HKLP}
we discovered that the eigenvalues $\{\lambda_i\}_i=1^{dN}$ (with $d$ being the space dimension) appear in two distinct families in generic amorphous solids, one corresponding
to eigenvalues of the hessian matrix that are only weakly sensitive to external strains; the other of
eigenvalues go to zero at certain values of the external strain, thus leading to a plastic failure. The first
group of modes is decently described by the the Debye model of an elastic body, but this is not the case for the second
group  corresponding to the density of plastic modes. 

For the purposes of the present letter it is enough to
write this excess part (in the thermodynamic limit) as $B(T)f_{\rm pl}\left(\frac{\lambda}{\langle \lambda_D \rangle}\right)$,
where the pre-factor $B(T)$ being strongly dependent on temperature. Particular models for this part were presented
in \cite{10KLP}. We do not need to specify a function here, and it is only important to understand that this function
is a partial characterization of the degree of disorder which grows upon approaching the glass transition.
Together with the standard Debye contribution one can approximate the low-frequency tail of the density of
states as
\begin{equation}
P(\lambda) \simeq A\left( \frac{\lambda}{\langle \lambda_D \rangle} \right)^{\frac{d-2}{2}}
+ B(T) f_{\rm pl}\left( \frac{\lambda}{\langle \lambda_D \rangle} \right) \ . \label{Poflam}
\end{equation}
Here $\lambda_D \simeq \mu \rho^{2/d - 1}$, is the Debye cutoff frequency and $\mu$ is the shear modulus.
The physical idea that allows the determination of the static typical scale is that the {\em minimal} eigenvalue
$\lambda_{\rm min}$ observed in a system of $N$ particles will be determined by either the first {\em or}
the second term in Eq. (\ref{Poflam}). For a system large enough, local disorder will be irrelevant in determining
$\lambda_{\rm min}$, and it will be decided by the Debye contribution. For small systems the opposite is true. Thus there exists a value of $N$ where a cross-over occurs. This cross-over is interpreted
in terms of a typical length-scale separating correlated disorder from asymptotic elasticity.

In Ref. \cite{11KLP} it was shown how to derive an implicit equation for $\frac{\langle \lambda_{\rm min}\rangle}
{\langle \lambda_{\rm D} \rangle}$ where the angular brackets represent an ensemble average over many systems prepared at the same temperature.
The equation takes the form
\begin{equation}
\frac{\langle \lambda_{\rm min} \rangle}{\langle \lambda_{\rm D} \rangle} =
{\C F}\left[ \xi_s^d(T)\left(\frac{1}{V} -
\frac{\tilde A d}{2} \left( \frac{\langle \lambda_{\rm min} \rangle }{\langle \lambda_{\rm D} \rangle}\right)^{d/2} \right) \right].
\label{ansatz}
\end{equation}
The typical scale $\xi_s(T)$
will be calculated by demanding that all the data calculated for different system sizes and temperatures should
collapse into a master curve just by appropriately choosing the $\xi_s(T)$.

As an example consider the analysis done for the glass forming Kob-Andersen binary mixture
\cite{95KA} at number density $\rho = 1.20$. The systems were equilibrated
at some temperature $T> T_g$ and then inherent structures
are calculated by direct energy minimization to the nearest local minimum of the potential
energy landscape. At this state
the Hessian was computed and the minimal eigenvalue was obtained using the Lanczos algorithm \cite{wik}. For a
given system size $N$ and temperature $T$ this procedure was repeated to have an average
$\langle \lambda_{\rm min}\rangle$ until convergence was achieved. At this point the temperature or the system
size were changed and the procedure was repeated, to eventually have a table of
$\langle \lambda_{\rm min}\rangle(N,T)$. In the left Panel of Fig.~\ref{lamminKA}, we have plotted the minimal eigenvalue rescaled by the characteristic Debye value as a function of system size $N$ for different temperatures. In the right Panel the same data is plotted
according to the scaling ansatz Eq.~(\ref{ansatz}) to extract the length-scale $\xi_s$ by collapsing the data
into a master curve. The resulting data collapse into a master curve indicates that our scaling ansatz
Eq.~(\ref{ansatz}) is obeyed to high precision. In the inset at the right panel we show how the typical scale
increases when the glass transition is approached.

We also showed in ~\cite{11KLP} how our typical scale helps in understanding other measures of disorder that
were proposed in the past. As an example we showed that the system size dependence of configurational entropy
$S_c(T)$ ~\cite{09KDS,SmarajitThesis2009} (the reader is referred to these
publications for a full description of the method and the results) of the Kob-Andersen model in 3 dimensions
can be explained using the typical scale proposed in ~\cite{11KLP}. This also gives us a direct
relation between structural relaxation time and the static
length-scale using the Adam-Gibbs relation ~\cite{65AG}. Below we will see how this relation along
with the static length-scale helps us to understand the effect of immobile particles on the dynamics
of supercooled liquid.

We start with the Adam-Gibbs relation Eq.~(\ref{AdamGibbs}). In the presence of immobile impurities
of density $\rho_{\rm im}$ the configurational
entropy changes, and we propose that it has a scaling form, reading
\begin{equation}
S_c(\rho_{\rm im},T)= S_c(T) g( \rho_{\rm im}\xi_s^d(T)) \ ,\label{Sc}
\end{equation}
where $g(x)$ is an unknown scaling function.
Thus with the presence of randomly pinned particles the relaxation time will depend on the  scaling function $g(x)$ and
$\rho_{im}\xi^d_s$ as
\begin{equation}
\tau_{\alpha}(\rho_{im},T) = \tau_0 \exp\left(\frac{A}{TS_c g(\rho_{im}\xi^d_s)}\right)\ .
\label{imAG}
\end{equation}
The scaling function $g(x)$ must have the following asymptotic behavior
\begin{equation}
g(x) \to 1, \quad \mbox{as} \,\, x \to 0.
\label{scalingFn}
\end{equation}
So using Eq.~(\ref{imAG}) and Eq.~(\ref{AdamGibbs}), we end up with following equality
\begin{equation}
\log\left[ \frac{\tau_{\alpha}(\rho_{im},T)}{\tau_{\alpha}(0,T)}\right] =
\frac{A}{TS_c }\left(\frac{1}{g(\rho_{im}\xi^d_s)} - 1\right)
\label{ratioRelx}
\end{equation}
Now using Eq.~(\ref{scalingFn}), we can expand the scaling function $g(x)$ for small $x$ as
\begin{equation}
g(x) \simeq 1 - |g^{'}_0| x + {\cal O}(x^2),
\label{appScalFn}
\end{equation}
We expect $g(x)$ to be a decreasing function of $x$ as we noticed before that with increasing disorder density
$\rho_{im}$ the relaxation time $\tau_{\alpha}$ increases so the first derivative calculated at $x=0$, $|g^{'}_0|$
will be negative. Putting Eq.~(\ref{appScalFn}) in Eq.~(\ref{ratioRelx}) we arrive at the following relation after
little bit of algebraic manipulation,
\begin{equation}
TS_c\log\left[ \frac{\tau_{\alpha}(\rho_{im},T)}{\tau_{\alpha}(0,T)}\right] \simeq
A\left( |g^{'}_0|\rho_{im}\xi^d_s  +  {\cal O} \left[(\rho_{im}\xi^d_s)^2\right] \right)
\label{appRatioRelx}
\end{equation}
This relation is expected to be valid only in the dilute disorder limit. Next we explain the details of the
experiment and the results that validate this scaling theory.

The experiment is done as follows. We first equilibrate a system of $N = 1000$ particles interacting via
the Kob-Andersen Potential \cite{95KA}. The interaction potential is given by
\begin{equation}
V_{\alpha\beta}(r)=4\epsilon_{\alpha\beta}[(\frac{\sigma_{\alpha\beta}}{r})^{12}-
(\frac{\sigma_{\alpha\beta}}{r})^{6}]
\end{equation}
where $\alpha,\beta \in \{A,B\}$ and $\epsilon_{AA}=1.0$, $\epsilon_{AB}=1.5$,
$\epsilon_{BB}=0.5$, $\sigma_{AA}=1.0$, $\sigma_{AB}=0.80$, $\sigma_{BB}=0.88$. The
Interaction Potential was cut off at 2.50$\sigma_{\alpha\beta}$. We have performed the simulations
at six different temperatures in range $T \in [{1.00,\, 0.45}]$.
Then we take the equilibrated system and randomly choose $N_{im} = N*\rho_{im}$ number of particles and freeze their
degrees of freedom and run the dynamics to calculate the relaxation time.

\begin{figure}
\includegraphics[scale = 0.40]{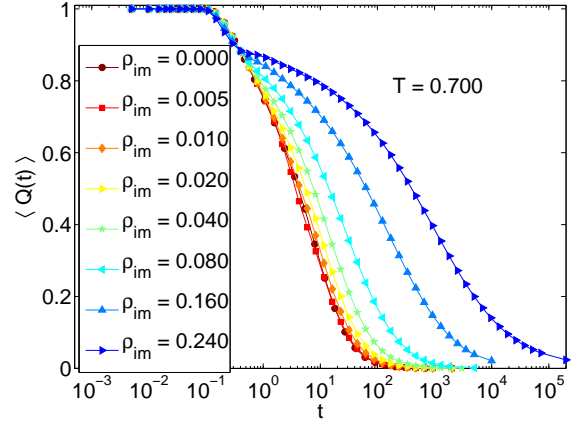}
\caption{Average Overlap function $\langle Q(t)\rangle$ is plotted for different disorder density $\rho_{im}$ for $T = 0.700$.}
\label{corlFn}
\end{figure}

We have calculated the overlap function defined below to estimate the structural relaxation time $\tau_{\alpha}$,
\begin{equation}
Q(t) = \frac{1}{N - N_{im}}\sum_{i=1}^{N\,\,\,\prime} w(|\vec{r}_i(t) - \vec{r}_i(0)|),
\end{equation}
where the weight function $w(x) = 1$ if $x<0.30$ and zero otherwise. The prime sign in the summation indicates
that immobile pinned particles are not included in the summation while calculating the correlation function.
 In Fig.~\ref{corlFn}, we plotted the
overlap function $Q(t)$ averaged over $50$ different realization of quenched disorder for different density
of disorder at temperature $T = 0.700$.
\begin{figure}
\includegraphics[scale = 0.450]{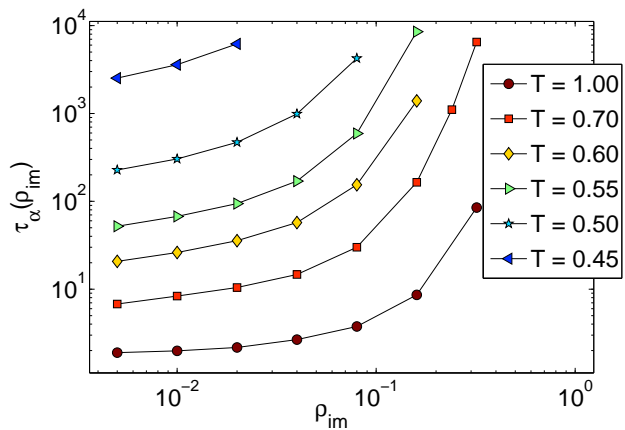}
\caption{The relaxation time $\tau_\alpha$ measured for all temperatures
plotted against density of pinned particles $\rho_{im}$.}
\label{tauImp}
\end{figure}
The Relaxation time $\tau_{\alpha}$ is defined to be the time where the correlation function $Q(t) = 1/e$, where $e$
is the base of natural logarithm.

In Fig.~\ref{tauImp}, we have shown the dependence of $\tau_{\alpha}$
as a function of disorder density $\rho_{im}$ for the six simulated temperatures. Note that as one decreases
the temperature the effect of disorder in the relaxation time kicks in for smaller disorder density. This can be
understood easily from the fact that the effect of the quenched disorder is to pin the correlated region to
one place causing hindrance to relaxation. Now as one decrease the temperature the size of these correlated
region grows and one needs less number of pinning sites to freeze the whole system.

\begin{figure}
\includegraphics[scale = 0.460]{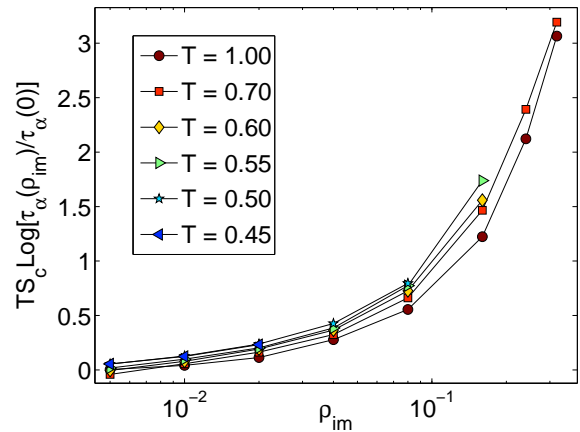}
\caption{The logarithm of the rescaled relaxation time $\tau_\alpha(\rho_{im})/\tau_{\alpha}(0)$ multiplied by
$TS_c$ measured for all the simulated temperatures plotted against $\rho_{im}$.}
\label{tauBeforeScaling}
\end{figure}
\begin{figure}[!h]
\includegraphics[scale = 0.50]{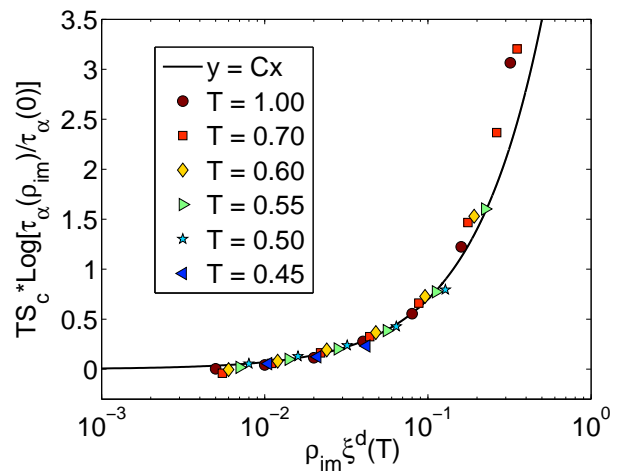}
\caption{The data collapse of relaxation time using the scaling ansatz \ref{ratioRelx}. The solid line is the
approximate scaling form derived in Eq.~\ref{appRatioRelx}.}
\label{tauScaling}
\end{figure}

In Fig.~\ref{tauBeforeScaling}, we have plotted $TS_c\log\left[ \tau_\alpha(\rho_{im})/\tau_{\alpha}(0)\right]$
measured for all the simulated temperatures against $\rho_{im}$ and in Fig.~\ref{tauScaling}, we have plotted
the same data but rescaling the abscissia by the appropriate power of static length-scale $\xi_s$. The
quality of the collapse suggests that indeed our proposed scaling ansatz Eq.~(\ref{ratioRelx}) is obeyed to a
high precission. The solid line is the approximate scaling form derived in Eq.~(\ref{appRatioRelx}). It is clear from
the figure that the approximate form of the scaling function is indeed a very good approximation in the dilute
disorder regime. We also want to point out here that Eq.~(\ref{appScalFn}) suggests that one can have an ideal
glass transition as a function of disorder density at a critical density of disorder $\rho^c_{im}(T)$ given by
\begin{equation}
\rho^c_{im}(T) \simeq \frac{1}{|g^{'}_0|\xi^d_s(T)}.
\end{equation}

In summary, we have showed that the dynamics of supercooled liquid under the influence of externally imposed
quenched disorder can be understood completely by our proposed static length-scale.
This analysis also suggests that one can get the static
length-scale by studying the dynamics of the supercooled liquids under externally imposed quenched disorder
instead of calculating it from the scaling of minimal eigenvalue which may be less trivial to access in experiments.
It would be really nice if one could experimentally study the dynamics of supercooled liquids under externally imposed
disorder, since the configurational entropy which is needed to have the data collapse is also easily calculable in experiments.
We hope that this kind of analysis will be done on other model glass formers and also in laboratory experiments
by other groups to enhance the understanding of the glass transition.

This work had been supported in part by an ERC ``ideas" grant, the Israel Science Foundation and by the German Israeli Foundation.
We benefitted from discussions with Eran Bouchbinder and Konrad Samwer.

\end{document}